\input epsf
\documentstyle[twocolumn,prl,aps]{revtex}
\begin{document}
\date{\today}
\draft
\title{UNWINDING SCALING VIOLATIONS IN PHASE ORDERING}
\author{A. D. Rutenberg and  A. J. Bray}
\address{Theoretical Physics Group, Department of Physics and Astronomy, 
The University of Manchester, M13 9PL, UK}
\maketitle
\begin{abstract}
The one-dimensional $O(2)$ model is the simplest example of a system with 
topological textures. The model exhibits anomalous 
ordering dynamics due to the appearance of {\em two} characteristic 
length scales: the phase {\em coherence} length, $L \sim t^{1/z}$, and 
the phase {\em winding} length, $L_{w} \sim L^{\chi}$. We derive the scaling 
law $z=2+\mu\chi$, where $\mu=0$ ($\mu=2$) for nonconserved (conserved) 
dynamics and $\chi=1/2$ for uncorrelated initial orientations. 
From hard-spin equations of motion, we consider the evolution of the 
topological defect density and recover a simple scaling description.
\end{abstract}
\pacs{05.70.Ln, 64.60.Cn, 64.60.My}


The scaling hypothesis
has played an important role in our understanding 
of the late-stage ordering dynamics of systems quenched from a homogeneous 
disordered phase into an ordered phase region with a broken symmetry
\cite{Furukawa85}. According to this hypothesis, the 
order parameter morphology at late times after the quench is statistically 
independent of time if all lengths are rescaled by a single 
characteristic length scale $L(t)$. This implies that the 
pair correlation function $C(r,t)$ of the order parameter should depend 
on its arguments only through the ratio $r/L(t)$. Recently, we have shown 
\cite{Bray94b}
that a natural extension of the scaling hypothesis to two-time
correlations, $C(r,t,t')=f(r/L(t),r/L(t'))$, 
supplemented by an understanding of the 
short-distance (i.e.\ $r \ll L(t)$) structure that follows from any 
singular topological defects  seeded by the quench \cite{Mermin79},
determines the 
late-time growth-law of $L(t)$. Any departure from these growth laws
implies a breakdown of the single-length scaling. 

Given the importance of the scaling phenomenology, it is important to 
look for exceptions to single-length scaling, and to try to understand 
them within a broader scaling framework. There  is 
evidence that some systems with nonsingular topological textures may 
violate conventional single-length scaling. 
Textures have a spatial extent, which can in principle introduce a new
characteristic length scale. 
The $O(n)$ model for an 
$n$-component vector field in spatial dimension $d=n-1$ provides a class 
of models with topological textures. Indeed, the 
$O(3)$ model in $d=2$
seems to have at least one of its characteristic scales growing as $t^{1/3}$ 
for nonconserved dynamics \cite{Bray90,Rutenberg94}, 
which contrasts with the anticipated $t^{1/2}$ 
growth for this system if scaling holds \cite{Bray94b,Rutenberg94}.

The $O(2)$ model (or `XY model') in spatial dimension $d=1$ is the 
simplest  system with topological textures. We show 
through the time-derivative correlations, $T(r,t) = \left.
\partial_t \partial_{t'} \right|_{t=t'} C(r,t,t')$, that 
single-length scaling fails in this system, both for 
conserved and non-conserved dynamics. The non-conserved 
case is exactly soluble \cite{Newman90,Rutenberg94b}: 
one finds that $C(r,t)$ scales with a length 
scale $t^{1/4}$, different from the $t^{1/2}$ scaling predicted from the 
scaling hypothesis \cite{Bray94b}, and $T(r,t)$ scales with
the same length-scale but with an anomalous time-dependent prefactor.
We show, within a general framework, that 
this discrepancy is due to the existence of {\em two} characteristic length 
scales, the `phase coherence length' $L \sim t^{1/2}$, and the `phase winding 
length', $L_w \sim t^{1/4}$.  Since $L_w$ is the typical
length scale over which the 
phase changes by order unity (see Figure 1), it provides the 
characteristic scale for the pair correlation function. 
The phase coherence length, $L$, drives the dynamics and enters into the
two-time correlation functions, such as  $T(r,t)$.

In contrast to the nonconserved model, the conserved model 
has not previously been addressed except by 
computer simulations. The same concepts are relevant to this case also, 
however, and by means of a simple scaling argument we find $L \sim t^{1/3}$ 
and $L_w \sim t^{1/6}$, the latter in agreement with recent simulations 
\cite{Rutenberg94b,Mondello93}. 

The issue of scaling in these systems is clarified by showing that 
the phase-difference correlation function $G(r,t)$ (\ref{EQ:G}), 
rather than the order parameter correlation function, exhibits 
a generalized  form of single-length scaling  with 
characteristic length $L$. The topological charge density, 
proportional to the phase gradient, provides an equivalent description 
that exemplifies the simple scaling description.

The key elements of this paper are i) the development of hard-spin
equations of motion for conserved dynamics, ii) the simplification of 
the equations of motion at late times using the separation of the length-scales
$L$ and $L_w$, iii) the scaling relations for the length-scales, and the
consequent forms for spin-spin correlations $C(r,t)$ and $T(r,t)$, and iv)
the unifying scaling description in terms of the topological charge density.

Figure 1 shows a typical configuration generated by a computer simulation 
with nonconserved dynamics. The configuration consists of sections of typical 
length $L$ where the order parameter winds in a given sense, alternating with 
antiwinding sections. The winding length $L_w$ is the typical distance 
between successive windings of $2 \pi$ in the phase. Since each complete 
winding (antiwinding) represents a topological 
texture (antitexture) in this system, $L_w$ is the characteristic texture 
size. During the phase ordering process, because isolated textures expand
\cite{Derrick64},
textures unwind by annihilating with adjacent 
antitextures at the boundaries between regions of positive and negative 
winding 

We begin with a heuristic argument relating $L$ and $L_w$.  
Consider a region of length $l$. If the phase angles in the initial 
condition have only short-range correlations, with correlation length $\xi_0$,  
then the initial net winding over the length $l$ is of order $(l/\xi_0)^{1/2}$. 
Because the total winding is a topological invariant, the net winding on 
scales much larger than the phase coherence length, $l \gg L$, 
will be unchanged.  At later times, the 
length $l$ contains of order $l/L$ sections, each winding in a given 
sense, with of order $L/L_w$ windings per section. The net winding is 
therefore $(l/\xi_0)^{1/2} \sim (L/L_w)(l/L)^{1/2}$, giving 
$L_w \sim (L\xi_0)^{1/2}$. (We assume here that the {\em fluctuations} 
in the total winding per section are comparable with the mean winding). 
This indicates that two time-dependent lengths characterize the system.

We now present an explicit calculation that verifies our heuristic argument
and gives the growth laws for $L$ and $L_w$. It is convenient to formulate 
the problem in terms of the $U(1)$ model for a complex scalar field 
$\phi = \rho \exp(i\theta)$. We take the conventional Ginzburg-Landau 
free-energy functional 
\begin{equation}
F[\phi] = \int dx\,[(\partial_x \phi)(\partial_x \phi^*) 
                    + (g/2)(1-\phi\phi^*)^2]\ .
\label{EQ:GL}
\end{equation}
The purely dissipative equation of motion is
\begin{equation}
\partial_t \phi = -(-\partial_x^2)^{\mu/2}\,\delta F/\delta\phi^*\ ,
\label{EQ:DYN}
\end{equation}
where $\mu=0$ and 2 for nonconserved and conserved dynamics respectively. 
It is mathematically convenient to take the limit $g \to \infty$, which 
imposes the constraint  $|\phi|=1$, corresponding to a non-linear sigma model 
or ``hard-spin''
description. This limit is taken by  writing $\phi=\exp(i\theta-\beta/g)$ in 
(\ref{EQ:DYN}), expanding in $\beta/g$, and retaining only terms of order 
unity. [ This method can be quite generally applied to conserved vector systems
by taking $\vec{\phi} = \exp(-\beta/g) \hat{\phi}$, where $|\hat{\phi}|=1$.]
For the 1D XY model, we obtain
\begin{equation}
i\dot{\theta}\exp(i\theta) = (-\partial_x^2)^{\mu/2}\,
                    [i\theta''-(\theta')^2 + 2\beta] \exp(i\theta)\ ,
\label{EQ:NLSM}
\end{equation}
where dots and primes indicate derivatives with respect to $t$ and $x$ 
respectively.  

Consider first the nonconserved case, $\mu=0$. Equating real and imaginary 
parts in (\ref{EQ:NLSM}) gives 
\begin{eqnarray}
\label{EQ:DIFF}
\dot{\theta} & = & \theta''\ , \\
\beta & = & (\theta')^2/2\ .
\label{EQ:SLAVED}
\end{eqnarray}
Thus the phase equation (\ref{EQ:DIFF}) decouples from the amplitude 
equation (\ref{EQ:SLAVED}), and the amplitude is slaved to the phase.  

For the conserved case, $\mu=2$, the same treatment yields coupled equations 
for $\theta$ and $\beta$. Motivated by Eq.\ (\ref{EQ:SLAVED}), we put 
$\beta = (\theta')^2/2 + \gamma$, where we anticipate that $\gamma$ will be 
negligible at late times.  The resulting 
equations are
\begin{eqnarray}
\label{EQ:PHASE}
\dot{\theta} & = & (\theta')^2\theta'' - \theta'''' -2\gamma\theta''
               - 4\gamma'\theta'\ ,\\
2\gamma'' & = & 2(\theta')^2\gamma + 2\theta'\theta''' + (\theta'')^2\ .
\label{EQ:AMP}
\end{eqnarray}
It is easy to show explicitly that these equations conserve the order 
parameter, i.e.\ $\partial_t \int dx\,\exp(i\theta) = 0$. 

As a consequence of the two length scales in the problem, the first term 
on the right of (\ref{EQ:PHASE}) dominates at late times, so that the phase 
equation (\ref{EQ:PHASE}) again decouples from the amplitude 
equation (\ref{EQ:AMP}) at late times. 
The key point is that while the typical {\em size} of $\theta'$ is given 
by $\theta' \sim 1/L_w$, the spatial {\em variation} of $\theta'$ occurs 
on the longer scale $L$ (see Figure 2). Thus each higher derivative 
generates an extra factor of $1/L$, giving $\theta'' \sim 1/LL_w$, 
$\theta''' \sim 1/L^2L_w$, etc. Thus on the right of (\ref{EQ:PHASE}) 
$\theta'''' \sim 1/L^3L_w$ is negligible compared to 
$(\theta')^2\theta'' \sim 1/LL_w^3$. Now look at Eq.\ (\ref{EQ:AMP}). 
Demanding that $(\theta')^2\gamma \sim \theta'\theta'''\sim (\theta'')^2 
\sim 1/L^2L_w^2$ gives $\gamma \sim 1/L^2$ (and on the left, $\gamma'' 
\sim 1/L^4$ is negligible). Putting this in (\ref{EQ:PHASE}), we find that 
the terms involving $\gamma$ are both of order $1/L^3L_w$ and therefore 
negligible at late times. Thus the first term on the right of (\ref{EQ:PHASE}) 
dominates at late times, giving the simplified dynamics 
\begin{equation}
\dot{\theta} = (\theta')^2\theta''\ .
\label{EQ:CONS}
\end{equation}
This equation is one of the central results of the paper, and represents 
a significant simplification of the original equation of motion. 
Although equation (\ref{EQ:CONS}) no longer conserves the order 
parameter at all times, the omitted terms on the right of (\ref{EQ:PHASE}) 
are of relative order $L_w^2/L^2 \sim 1/L$, and the conservation is 
asymptotically recovered  at late times. 

The key step in deriving the growth exponents for $L$ and $L_w$ is 
to transform from the phase variable $\theta$ to the phase gradient 
$y \equiv \theta'$. Note that 
$q(x) \equiv y(x)/2\pi$ is just the local winding rate or the 
`topological charge density' at point $x$. Our basic assumption is that,
whereas the order parameter representation 
sketched in Figure 1 can never be made scale invariant due to the 
two different length scales, the same morphology in the $y$-representation 
of Figure 2 {\em is} scale invariant under a simultaneous rescaling of 
$x$ by  $L$ and $y$ by $1/L_w \sim 1/L^\chi$ (where we anticipate $\chi=1/2$ 
from our heuristic argument).  This is an important generalization
of the standard dynamical scaling hypothesis, and is confirmed by exact
calculation for non-conserved dynamics and by simulation
for conserved dynamics \cite{Rutenberg94b}.

For compactness, we combine (\ref{EQ:DIFF}) and (\ref{EQ:CONS}) as the single 
equation $\dot{\theta} = (\theta'^2)^{\mu/2}\theta''$. In terms of $y$, this 
reads
$\dot{y} = [(y^2)^{\mu/2} y']'\ ,\ \ \ \ \ (y \equiv \theta')$.
Making the scale transformations $x \to bx$, $t \to b^zt$, $y \to b^{-\chi}y$ 
and demanding scale invariant behavior, gives our 
main result:  
\begin{equation}
z = 2 + \mu\chi\ .
\label{EQ:z}
\end{equation}

To determine the exponent $\chi$, and to exemplify the scaling, it is 
convenient to introduce the squared phase difference correlation function
\begin{eqnarray}
G(r,t) & = & \langle [\theta(x+r,t)-\theta(x,t)]^2 \rangle\ \nonumber \\
       & = & L^{2(1-\chi)}\,g(r/L)\ ,
\label{EQ:G}
\end{eqnarray}
where the scaling form follows from the scaling transformations above
and from noting that phase differences scale as 
$L^{1-\chi}$. Alternatively, we could work with the phase gradient (or 
topological charge 
density) correlation function $H(r,t) = \langle y(x+r)y(x) \rangle 
= L^{-2\chi} h(r/L)$. 

The angled brackets in (\ref{EQ:G}) represent an average over an ensemble of 
initial conditions. A natural choice of initial conditions is the gaussian 
distribution
\begin{equation}
\label{EQ:IC}
P[\theta(x,0)] \propto \exp\{ -\sum_k \theta_k(0)\theta_{-k}(0)/2\sigma_k\}\ ,
\end{equation}
where $\theta_k(0)$ is the Fourier amplitude of $\theta(x,0)$. The pair 
correlation function for the order parameter at $t=0$ then takes the form 
$ C(r,0)  =  \langle \phi(x+r,0) \phi^*(x,0) \rangle 
        =  \exp\{-\langle[\theta(x+r,0)-\theta(x,0)]^2\rangle/2\}
	    =  \exp\{-\sum_k \sigma_k(1-\cos kr)\}$ .
Choosing $\sigma_k = 2/\xi_0 k^2$ yields $C(r,0)=\exp(-r/\xi_0)$, appropriate 
to a quench from a disordered phase with correlation length $\xi_0$. This 
corresponds to a `random walk' of the initial phase angles, so that
$G(r,0) = 2r/\xi_0$.

Now consider the dynamics. Since the topological charge is locally conserved,
the development of phase coherence at scale $L$ due to texture-antitexture 
annihilation does not affect the phase-difference correlation function at 
larger separations $r \gg L$, i.e.\ $G(r,t) \to 2r/\xi_0$ in this limit. 
Hence from Eq. (\ref{EQ:G}), the scaling function 
$g(x) \sim x$ for $x \to \infty$, and since $L$ must drop out 
in this limit we have 
\begin{equation}
\chi=1/2\ .
\end{equation}
Putting this into (\ref{EQ:z}) gives 
$z=2+\mu/2$, and so 
\begin{eqnarray}
\label{EQ:GROWTHL}
L & \sim & t^{2/(4+\mu)},  \\
\label{EQ:GROWTHLW}
L_w & \sim & L^{1/2} \sim t^{1/(4+\mu)}\ . 
\end{eqnarray}

Previous studies of phase ordering systems have usually concentrated on the 
pair correlation function $C(r,t)$, and its Fourier transform, the structure 
factor $S(k,t)$. In the present context, the phase difference correlation 
function $G(r,t)$ is more appropriate, as it reveals the full scaling structure 
(\ref{EQ:G}). The scaling properties of $C(r,t)$ can, however, be inferred. 
The usual pair correlation function is given by  
$ C(r,t)  =  \langle \exp\{i[\theta(x+r,t) - \theta(x,t)]\} \rangle 
        =  \langle \exp\{i(ry + r^2y'/2 + r^3y''/6 +\cdots\} \rangle$,
where the second equality follows from the Taylor series expansion of 
$\theta(x+r,t)$. In the late-time 
limit $r \to \infty$, $L_w \to \infty$, with $r/L_w$ 
fixed, only the leading term in the expansion survives, because 
$ry \sim r/L_w$ is of order unity, while $r^2y' \sim r^2/LL_w$ is of order 
$L_w/L \ll 1$, and the higher terms are smaller still. This limit 
probes correlations on the scale $L_w$, since $L_w^{-1}$ sets the scale
of $y \equiv \theta'$, so that
\begin{equation}
C(r,t)  = \langle \exp(iry) \rangle  = f(r/L_w) , 
\label{EQ:C}
\end{equation}
where $L_w \sim t^{1/4}$ for $\mu=0$ and $t^{1/6}$ for $\mu=2$. 
Because the structure factor $S(k,t)$ 
is the spatial Fourier transform of $C(r,t)$, 
from (\ref{EQ:C}) we see that $S(k,t) = P(k,t)$, where $P(y,t)$ is the 
single-point probability distribution for $y$. 
For $\mu=0$, the linear 
dynamics (\ref{EQ:DIFF}) combined with the gaussian initial condition 
(\ref{EQ:IC}) ensures that the probability distribution, and hence $S(k,t)$ 
and $C(r,t)$ are gaussian at all times. The exact solution of the model 
\cite{Newman90,Rutenberg94b} 
confirms this feature, with the expected scale length $L_w \sim t^{1/4}$. 
For the conserved case, the conservation requires that $S(k,t)$ vanish at 
$k=0$, implying $P(0,t) = 0$ in the scaling limit. Thus $P(y,t)$ {\em cannot} 
be gaussian in this case, but must have a double peaked structure. Numerical 
studies \cite{Rutenberg94b,Mondello93} are consistent with the result 
$L_w \sim t^{1/6}$ derived above,  
and indicate \cite{Rutenberg94b} that $P(y,t)$ is approximately 
described by the form $P(y,t) \sim L_w^3y^2\exp(-{\text const}\; y^2L_w^2)$. 

Time-derivative correlations probe the scaling properties of the full two-time 
correlations $C(r,t,t')$. In the same limit $r \to \infty$, 
$L_w \to \infty$, with $r/L_w$ fixed, we have 
$T(r,t) \equiv \partial_t \partial_{t'} |_{t=t'} C(r,t,t') = 
\langle \dot{\theta}^2 \exp\{i[\theta(x+r,t) - \theta(x,t)]\} \rangle
= \langle y'^2 y^{2 \mu} \exp(iry) \rangle$,
where we have used equations (\ref{EQ:DIFF}) and (\ref{EQ:CONS}) for
$\dot{\theta}$. For the non-conserved case, because the variables are
gaussian and $\langle y y' \rangle = \langle (y^2)' \rangle /2 = 0$,
then $T(r,t) = \langle y'^2 \rangle \langle \exp (iry) \rangle =
\langle y'^2 \rangle C(r,t)$. For the conserved case, the phase variables
are not gaussian so $T(r,t)$ is not simply proportional to $C(r,t)$. In both
cases, we use $y \sim L_w^{-1}$ and $y' \sim (L L_w)^{-1}$ and the
growth laws of equations (\ref{EQ:GROWTHL}) and (\ref{EQ:GROWTHLW}) to determine
\begin{equation}
\label{EQ:TR}
	T(r,t) = t^{-2(\mu+3)/(\mu+4)} \tilde{f}(r/L_w),
\end{equation}
which breaks dynamical scaling because the time-dependent
amplitude is not proportional to $t^{-2}$ \cite{Bray94b}. 
Because the phase dynamics
involves spatial gradients of $y$, the second length-scale $L$ is 
introduced and dynamical scaling is broken.

These results can be generalized to a broader class of 
correlated initial conditions which includes a conventional scaling solution.
If we  take $\sigma_k \sim k^{-\alpha}$ in (\ref{EQ:IC}),we obtain 
$G(r,0) \sim r^{\alpha-1}$, provided $1<\alpha<3$. 
The requirement due to local phase conservation 
that this form be recovered from the general scaling 
form (\ref{EQ:G}) when $r \gg L$ fixes 
\begin{eqnarray}
\label{EQ:chi}
\chi   =  (3-\alpha)/2\ \ \  ;\ \ \ \ z   = 2 + \mu(3-\alpha)/2\ ,
\end{eqnarray}
where we have applied Eq.\ (\ref{EQ:z}).
For $\mu=0$ we still have $z=2$ with $L \sim t^{1/2}$ and 
$L_w \sim t^{(3-\alpha)/4}$, but for $\mu=2$  we obtain $z=5-\alpha$,
giving $L \sim t^{1/(5-\alpha)}$ and $L_w \sim L^{(3-\alpha)/2(5-\alpha)}$. 
For $\alpha=1$, we obtain $G(r,0) \sim \ln r$, implying a power-law 
decay of $C(r,0)$.
Simple scaling is recovered in the limit $\alpha \to 1$ since 
$\chi \to 1$ implies that 
$L_w$ and $L$ both grow in the same way,
 with characteristic scale $L \sim t^{1/2}$ for $\mu=0$, and 
$L \sim t^{1/4}$ for $\mu=2$. These growth laws are just what we expect for 
a one-dimensional $O(2)$ system with simple scaling \cite{Bray94b}. 

The essence of the `energy scaling' 
approach that determines the growth laws for 
single-length scaling \cite{Bray94b} 
can be used for an alternative derivation of 
the central result (\ref{EQ:z}). In the hard-spin limit, the 
free-energy functional (\ref{EQ:GL}) becomes $F[\theta]=\int dx\,\theta'^2$. 
The energy density is therefore 
$\epsilon = \langle \theta'^2 \rangle \sim L_w^{-2} \sim L^{-2\chi}$. 
This gives the energy density dissipation rate as 
$\dot{\epsilon} \sim -\dot{L}L^{-(2\chi+1)} $.
However, $\dot{\epsilon}$ may be independently estimated via 
$\dot{\epsilon} = \langle (\delta F/\delta \theta)\dot{\theta} \rangle  
\sim -\langle \theta''^2 (\theta'^2)^{\mu/2}\rangle 
\sim -L_w^{-(2+\mu)}L^{-2} \sim -L^{-[2+\chi(2+\mu)]}$.
Equating these two estimates gives $L \sim t^{1/(2+\mu\chi)}$.

To summarize, we have shown that the ordering dynamics of the $O(2)$ model 
in $d=1$ involves two characteristic length scales: $L_w \sim t^{1/(4+\mu)}$ 
which 
acts as the scaling length for the order parameter correlation functions, 
though $T(r,t)$, and by implication $C(r,t,t')$,
do not satisfy standard dynamical scaling,
and $L \sim t^{2/(4+\mu)}$ which is the scaling length for correlations of the 
phase difference (or phase gradient).  
Working with the phase 
gradient, which is proportional to the topological charge density for this
system,  is necessary to provide a unifying framework. 
It is only by considering correlations of the phase differences that a 
simpler scaling description emerges.
It will be interesting to see to what extent scaling violations, and/or
a simplified description in terms of the topological charge density,
occur in higher-dimensional texture systems \cite{Rutenberg94}. 

\acknowledgments

We thank T. Blum, W. Zakrzewski, and M. Zapotocky for discussions, and the
Isaac Newton Institute for hospitality.

\begin{figure}[h]
\begin{center}
\mbox{
\epsfxsize=3.0in
\epsfbox{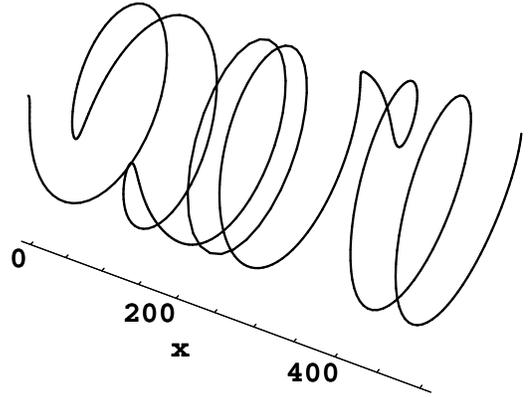} }
\end{center}
\caption{A section of a system from a simulation using 
non-conserved dynamics. Distance
along the system is shown by the scale and  the unit-magnitude 
order parameter is shown in the orthogonal plane. 
The windings of $ \pm 2 \pi$ (textures/anti-textures)
of scale $L_w$ and the clusters of monotonic winding of a larger 
scale $L$ are evident.
}
\label{FIG:HELIX}
\end{figure}

\begin{figure}[h]
\begin{center}
\mbox{
\epsfxsize=3.0in
\epsfbox{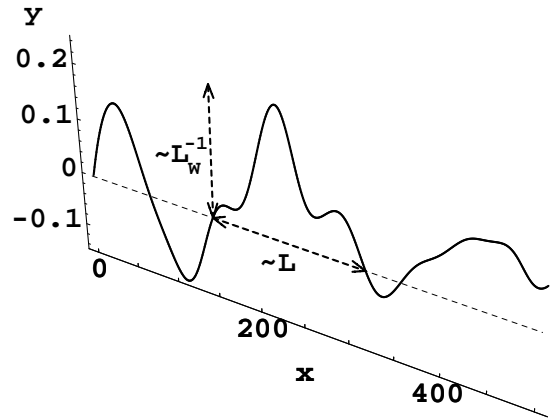} }
\end{center}
\caption{The phase gradient $y \equiv \theta'$ vs. distance 
is shown corresponding
to the section of system shown in Figure 1. The characteristic length
scales are indicated: $L$ is the average distance between zeros of $\theta'$,
while $2 \pi/ L_w$ is the characteristic magnitude of $\theta'$.
}
\label{FIG:DTHETA}
\end{figure}

\end{document}